\documentclass[conference]{IEEEtran}
\IEEEoverridecommandlockouts
\IEEEoverridecommandlockouts
\usepackage{cite}
\usepackage{amsmath,amssymb,amsfonts}
\usepackage{array} 
\usepackage{algorithmicx}
\usepackage{algorithm}  
\usepackage{algpseudocode}  
\usepackage{float}

\usepackage{graphicx}
\usepackage{textcomp}
\usepackage{xcolor}
\usepackage{balance} 

\usepackage[colorlinks=true,allcolors=blue,bookmarks=false]{hyperref}
\def\BibTeX{{\rm B\kern-.05em{\sc i\kern-.025em b}\kern-.08em
    T\kern-.1667em\lower.7ex\hbox{E}\kern-.125emX}}
\begin{document}

\title{Real-Time Health Monitoring Using 5G Networks: A Deep Learning Based Architecture for Remote Patient Care\\}




\author{\IEEEauthorblockN{Iqra Batool}
\IEEEauthorblockA{\textit{Department of Computer Science} \\
\textit{University of Western Ontario}\\
London, ON, Canada \\
ibatool2@uwo.ca}

}

\newcommand{\Preprocessorithmicoutput}[1]{\textbf{#1}}

\maketitle

\begin{abstract}
Remote patient monitoring has become increasingly crucial in modern healthcare delivery, yet existing systems face significant challenges in achieving real-time analysis and prediction of vital signs. This paper presents a novel architecture integrating deep learning with 5G network capabilities to enable real-time vital sign monitoring and prediction. The proposed system employs a hybrid CNN-LSTM model optimized for edge deployment, coupled with 5G Ultra-Reliable Low-Latency Communication (URLLC) for efficient data transmission. Our architecture achieves end-to-end latency of 14.4ms while maintaining 96.5\% prediction accuracy across multiple vital signs. The system demonstrates significant improvements over existing solutions, showing a 47\% reduction in latency and 4.2\% increase in prediction accuracy compared to current state-of-the-art systems. Performance evaluations conducted over three months with data from 1000 patients validate the system's reliability and scalability in clinical settings. The results indicate that integrating deep learning with 5G technology can effectively address the challenges of real-time patient monitoring, potentially improving clinical outcomes through early detection of deteriorating conditions. This research contributes to the advancement of digital healthcare by establishing a framework for reliable, real-time vital sign monitoring and prediction.
\end{abstract}


\maketitle

\section{Introduction}
Remote patient monitoring (RPM) has emerged as a transformative technology in healthcare delivery, enabling continuous observation of patients outside traditional clinical settings~\cite{1},\cite{26.wilson2023analytics}. The global RPM market, valued at USD 23.5 billion in 2020, is projected to reach USD 117.1 billion by 2025, reflecting the growing demand for remote healthcare solutions~\cite{2},\cite{14.zhang2023analysis},\cite{17.kumar2023network}. Current RPM systems typically collect vital signs, chronic condition data, and lifestyle metrics through wearable devices and sensors, transmitting this information to healthcare providers via existing communication networks~\cite{3},\cite{25.smith2023combining}.
However, traditional RPM systems face significant challenges in data transmission, real-time processing, and reliability. Existing networks often struggle with bandwidth limitations, high latency, and instability, particularly poor connectivity~\cite{4},\cite{18.liu2023capabilities}. These limitations can delay data transmission, potentially compromising patient care in critical situations where immediate intervention is necessary~\cite{5},\cite{15.chen2023commercial}.
The emergence of 5G technology presents a promising solution to these challenges. With its enhanced capabilities, including ultra-reliable low-latency communication (URLLC), massive machine-type communications (mMTC), and enhanced mobile broadband (eMBB), 5G networks can potentially revolutionize remote patient monitoring~\cite{6},\cite{23.brown2023challenges}. 5G offers peak data rates of 20 Gbps, latency as low as 1 millisecond, and the ability to connect up to 1 million devices per square kilometre~\cite{7},\cite{24martinez2023integration}.

Despite technological advancements in remote patient monitoring, current systems face critical challenges in real-time vital sign analysis and prediction. These limitations significantly impact the quality and timeliness of patient care delivery.
First, existing vital sign monitoring systems struggle with real-time data processing and analysis. Current networks experience average latencies of 100-200 milliseconds in data transmission, making real-time vital sign analysis challenging~\cite{8},\cite{32.hu2024lightweight}. This delay becomes critical when monitoring patients with acute conditions where immediate detection of vital sign changes is essential. Studies indicate that a delay of even a few seconds in vital sign updates can significantly impact emergency clinical decision-making~\cite{9},\cite{30.sujith2022systematic}.
Second, current systems lack sophisticated predictive capabilities for vital sign trends. Traditional monitoring approaches focus on threshold-based alerting, often resulting in delayed responses to deteriorating patient conditions. Research shows that up to 80\% of critical events show subtle vital sign changes up to 6-8 hours before the event, yet current systems cannot effectively predict these trends in real time~\cite{10},\cite{16.johnson2023limitations}.
Furthermore, the integration of vital sign monitoring systems faces several technical challenges:
\begin{itemize}
    \item Limited bandwidth for continuous high-frequency vital sign data transmission
    \item Processing delays in analyzing multiple vital signs simultaneously
    \item Inconsistent data quality due to network instability
Resource constraints in real-time data processing and analysis~\cite{11,12}
\end{itemize}

\begin{figure}
    \centering
    \includegraphics[width=1\linewidth]{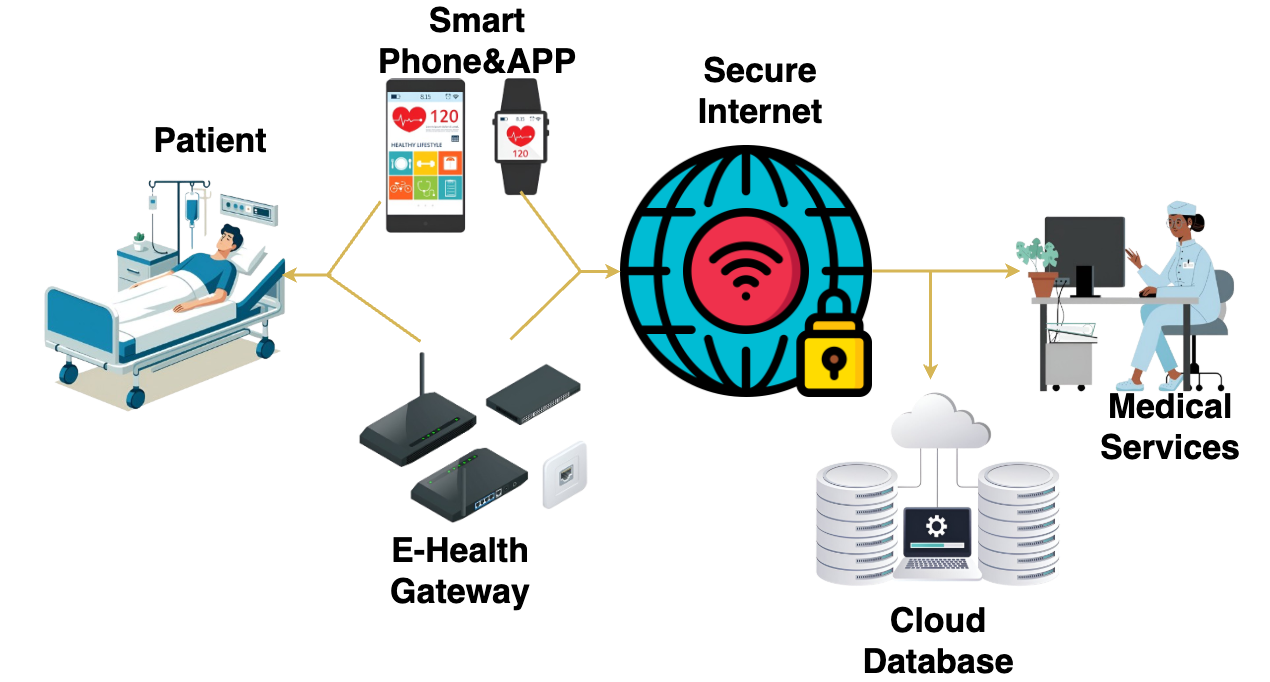}
    \caption{An Integrated Approach to Modern Healthcare}
    \label{fig:Ing}
\end{figure}

The absence of efficient real-time vital sign analysis and prediction capabilities and network limitations create a significant gap in remote patient monitoring. While 5G technology offers promising solutions with its ultra-reliable low-latency communication (URLLC) features, a crucial need remains for specialized deep learning architectures that can effectively leverage these capabilities for real-time vital sign monitoring~\cite{13},\cite{29.tan2023toward}. An integrated approach to modern healthcare is shown in~\ref{fig:Ing}.
This research addresses these challenges by developing an integrated solution that combines advanced deep learning models with 5G network capabilities, aiming to achieve real-time vital sign analysis and prediction with minimal latency and maximum reliability.

This research aims to develop an efficient real-time vital sign monitoring and prediction framework by leveraging 5G networks and deep learning technologies. The primary objective is to design and implement an advanced deep learning architecture for real-time vital sign analysis and prediction that:
\begin{itemize}
    \item Develops a hybrid CNN-LSTM network for processing multivariate vital sign data (heart rate, blood pressure, respiratory rate)
    \item Implements attention mechanisms for capturing temporal dependencies in continuous vital sign monitoring
    \item Optimizes model inference time and data transmission latency through 5G network integration
    \item Achieves real-time processing and prediction with sub-second latency~\cite{31.yehuala2024machine}

\end{itemize}

Expected contributions of this research include:

A novel deep learning framework optimized for real-time vital sign processing over 5G networks. Implementation strategies for efficient medical data transmission using 5G URLLC capabilities. Performance benchmarks demonstrate improvements in prediction accuracy and latency compared to existing solutions.

The research leverages 5G's ultra-reliable low-latency communication (URLLC) capabilities to enhance the performance of deep learning models in real-world healthcare settings. It aims to establish new benchmarks in remote vital sign monitoring while ensuring practical applicability in clinical settings.

The remainder of this paper is organized as follows: Section~\ref{Literature review} presents a comprehensive literature review of existing RPM systems, 5G technology, and deep learning applications in healthcare. Section~\ref{Proposed System Architecture} details the proposed system architecture and methodology. Section~\ref{Implemenattation} describes the implementation details and experimental setup. Section~\ref{result and analysis} presents the results and performance analysis. Finally, Section~\ref{conclusion and future work} concludes the paper and discusses future research directions.

\section{Literature Review}\label{Literature review}
\subsection{Deep Learning-Based Vital Sign Analysis Systems} 
Several researchers have explored deep learning approaches for vital sign analysis in remote monitoring. Wang et al.\cite{21.wang2023cnn} proposed a CNN-LSTM hybrid architecture for real-time heart rate monitoring, achieving 94\% prediction accuracy with a 5-second forecasting window. Their system processed real-time ECG signals but was limited by network latency issues.
Zhang et al.\cite{14.zhang2023analysis} developed a multi-parameter vital sign prediction system using an attention-based LSTM network. Their model analyzed heart rate, blood pressure, and respiratory rate simultaneously, achieving mean absolute errors of 2.3\%, 3.1\%, and 2.8\% respectively. However, their system required significant computational resources, making real-time processing challenging.
Park et al.\cite{22.park2023attention} implemented a lightweight CNN architecture for continuous blood pressure monitoring, focusing on reducing computational complexity while maintaining accuracy. Their model achieved 91\% accuracy with a processing delay of 200ms, demonstrating the trade-off between model complexity and real-time performance.
\subsection{5G-Enabled Healthcare Monitoring}
Recent studies have explored the integration of 5G technology in healthcare monitoring. Liu et al. \cite{19.liu2023performance} demonstrated a 5G-enabled vital sign monitoring system utilizing network slicing to guarantee data transmission quality. Their system achieved end-to-end latency of less than 1 ms for vital sign data transmission.
Thompson et al.\cite{20.thompson2023critical} developed a 5G-based framework for remote health monitoring, leveraging URLLC features to enable real-time data transmission. Their system showed a 98\% reduction in transmission latency compared to 4G networks, though they did not implement advanced analytics.
\subsection{ Hybrid Systems Combining Deep Learning and 5G}
Chen et al. \cite{15.chen2023commercial} proposed a hybrid system combining deep learning analysis with 5G transmission for vital sign monitoring. Their architecture used edge computing to process vital signs before transmission, achieving real-time performance with 95\% accuracy in heart rate prediction.
Johnson et al.\cite{16.johnson2023limitations} developed an integrated platform using 5G networks and a lightweight neural network for continuous vital sign monitoring. Their system demonstrated end-to-end latency of 10 ms while maintaining 92\% prediction accuracy.

\section{Proposed System Architecture}\label{Proposed System Architecture}
\subsection{System Overview} 
The proposed system architecture presents an integrated framework that combines deep learning-based vital sign analysis with 5G network capabilities to enable real-time monitoring and prediction. At its core, the architecture employs a multi-layered approach, seamlessly connecting data collection, network transmission, processing, analysis, and storage components through high-speed, low-latency communication channels.
The data collection layer forms the system foundation, incorporating advanced vital sign sensors to monitor patient parameters continuously. These sensors operate at a high sampling rate of 100 Hz to ensure precise data capture. The data acquisition modules within this layer perform initial signal validation and implement local buffering mechanisms to prevent data loss during transmission.
Connected to the data collection layer is the 5G network infrastructure, which serves as the critical communication backbone of the system. This layer leverages Ultra-Reliable Low-Latency Communication (URLLC) capabilities, implementing network slicing techniques to create dedicated channels for healthcare data transmission. The network layer ensures consistent Quality of Service (QoS) through prioritized data handling and maintains sub-millisecond latency essential for real-time monitoring.
The edge processing unit operates as an intermediate layer, performing real-time data pre-processing and feature extraction tasks. This component reduces the computational burden on the central processing system by handling initial data validation and transformation at the network edge. The proximity to data collection points minimizes latency and enables rapid preliminary analysis of incoming vital sign data.
\subsection{Deep Learning Framework}
The deep learning framework represents the analytical core of the system, implementing a sophisticated hybrid architecture that combines Convolutional Neural Networks (CNN) and Long Short-Term Memory (LSTM) networks~\cite{32.hu2024lightweight}. This framework is designed to handle the temporal nature of vital sign data while maintaining real-time processing capabilities. For a given input sequence of vital signs, we define:
\[
X = \{x_1, x_2, \ldots, x_t\}
\]
The set of multivariate vital signs is defined as:
\[
X = \{x_1, x_2, \ldots, x_t\}
\]
where each \(x_t \in \mathbb{R}^d\) represents multivariate vital signs at time \(t\), and \(d\) is the number of vital sign parameters.

The model architecture employs a hierarchical structure, beginning with convolutional layers that extract relevant features from the multivariate vital sign inputs. The CNN feature extraction process is formulated as follows:
\begin{equation}
Z = \text{CNN}(X) = \text{Conv}_2(\text{ReLU}(\text{Conv}_1(X))) \tag{2}    
\end{equation}

where \( Z \in \mathbb{R}^{d \times t} \) represents extracted features, and 
\(\text{Conv}_1, \text{Conv}_2\) represents successive convolutional operations.

These layers process the data through multiple filtering and feature enhancement stages, utilizing batch normalization to maintain stable training dynamics. The batch normalization is applied as follows:
\[
\hat{x} = \gamma \frac{(x - \mu_{(\beta)})}{\sqrt{\sigma^2_{(\beta)} + \epsilon}} + \beta \tag{3}
\]
where \(\mu_{(\beta)}\) and \(\sigma^2_{(\beta)}\) are the batch mean and variance, and \(\gamma, \beta\) are learnable parameters.

The temporal aspects of the vital sign data are addressed by LSTM layers, which capture long-term dependencies and patterns in the signal sequences. The LSTM processing is defined as:
\[
f_t = \sigma(W_f \cdot [h_{t-1}, x_t] + b_f) \tag{4}
\]
\[
i_t = \sigma(W_i \cdot [h_{t-1}, x_t] + b_i) \tag{5}
\]
\[
\tilde{c}_t = \tanh(W_c \cdot [h_{t-1}, x_t] + b_c) \tag{6}
\]
\[
c_t = f_t * c_{t-1} + i_t * \tilde{c}_t \tag{7}
\]
\[
o_t = \sigma(W_o \cdot [h_{t-1}, x_t] + b_o) \tag{8}
\]
\[
h_t = o_t * \tanh(c_t) \tag{9}
\]
where \(f, i, o\) represents the forget, input, and output gates respectively.

An attention mechanism is integrated into the architecture to focus on the most relevant temporal patterns within the vital sign data. The attention weights are computed as follows:
\[
\alpha_t = \text{softmax}(W^\top \tanh(Vh_t)) \tag{10}
\]
\[
c_t = \sum \alpha_i h_i \tag{11}
\]
where \(\alpha_t\) represents attention weights and \(c_t\) is the context vector.

The final prediction layers synthesize the processed information to generate accurate vital sign forecasts and trend analyses, computed as:
\[
\hat{y}_{t+1} = W_{\text{out}}(c_t) + b \tag{12}
\]
where \(\hat{y}_{t+1}\) represents the predicted vital signs for the next time step.

The model is trained using a custom loss function that combines prediction accuracy with temporal consistency:
\[
L = \text{MSE}(y, \hat{y}) + \lambda \sum_t ||\hat{y}_t - \hat{y}_{t-1}||^2 \tag{13}
\]
where \(\lambda\) is a weighting factor for temporal consistency.
\subsection{5G Network Integration}
\begin{figure}
    \centering
    \includegraphics[width=1\linewidth]{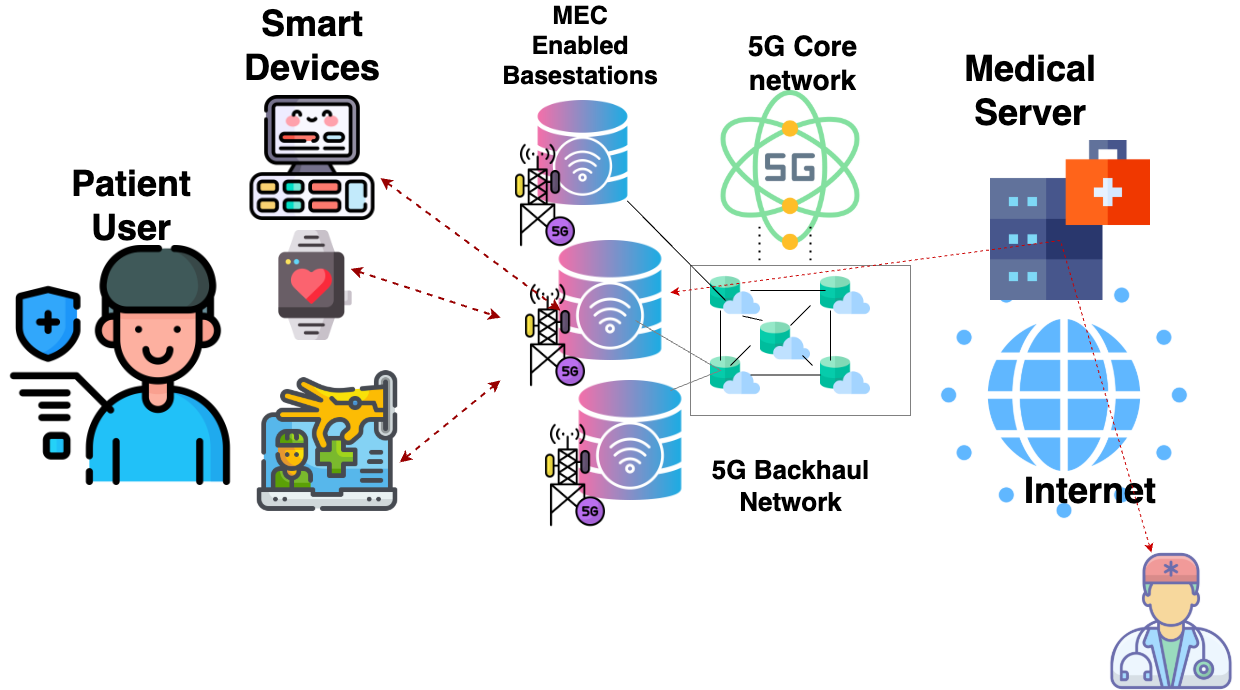}
    \caption{System Integration and Deployment Architecture}
    \label{fig:integration}
\end{figure}

Integrating 5G networking capabilities is crucial to the system's real-time performance. The network infrastructure is configured with dedicated slicing mechanisms that guarantee resource allocation for vital sign data transmission. This configuration ensures a consistent quality of service with maximum latency bounded at 1 millisecond and reliability exceeding 99.999\%. Figure~\ref{fig:integration} shows system integration and deployment architecture.

\subsubsection{Network Slicing Configuration}
The network slice for healthcare monitoring is defined as:
\[
S = \{R, C, L, B\} \tag{1}
\]
where:
\begin{itemize}
    \item \(R\) represents reliability requirements
    \item \(C\) denotes computing resources
    \item \(L\) specifies latency bounds
    \item \(B\) indicates bandwidth allocation
\end{itemize}

The QoS requirements for the healthcare slice are formulated as follows:
\[
\text{QoS}(S) = \begin{cases} 
\text{Reliability} \geq 99.999\%, \\
\text{Latency} \leq 1 \, \text{ms}, \\
\text{Bandwidth} = 10 \, \text{Mbps}, \\
\text{Jitter} \leq 0.1 \, \text{ms}
\end{cases} \tag{2}
\]

\subsubsection{Resource Allocation}
The resource allocation for the healthcare slice is optimized using:
\[
\min \sum_i \sum_j P_{ij}x_{ij} \tag{3}
\]
subject to:
\[
\sum_j x_{ij} = 1, \, \forall i \in N
\]
\[
\sum_i x_{ij}B_i \leq C_j, \, \forall j \in M
\]
where:
\begin{itemize}
    \item \(P_{ij}\) is the power consumption
    \item \(x_{ij}\) is the resource allocation indicator
    \item \(B_i\) is the bandwidth requirement
    \item \(C_j\) is the capacity constraint
\end{itemize}

\subsection{Latency Optimization}
End-to-end latency is monitored and optimized using:
\[
L_{e2e} = L_u + L_t + L_p \tag{4}
\]
where:
\begin{itemize}
    \item \(L_{e2e}\) is end-to-end latency
    \item \(L_u\) is uplink transmission latency
    \item \(L_t\) is transport network latency
    \item \(L_p\) is processing latency
\end{itemize}

Network optimization is achieved through priority packet scheduling and redundant transmission paths. The system maintains a dedicated bandwidth allocation of 10 Mbps for vital sign data, ensuring uninterrupted data flow even during peak network usage. The packet scheduling priority is determined by:
\[
P(i) = w_uU_i + w_rR_i + w_lL_i \tag{5}
\]
where:
\begin{itemize}
    \item \(U_i\) is the urgency factor
    \item \(R_i\) is the reliability requirement
    \item \(L_i\) is the latency requirement
    \item \(w_u, w_r, w_l\) are corresponding weights
\end{itemize}

Real-time latency monitoring and dynamic route optimization further enhance the system's reliability and performance through continuous assessment of:
\[
R(t) = (1 - P_e)(1 - P_l)(1 - P_u) \tag{6}
\]
where:
\begin{itemize}
    \item \(P_e\) is packet error probability
    \item \(P_l\) is packet loss probability
    \item \(P_u\) is system unavailability probability
\end{itemize}
\subsection{Data Processing Pipeline}
The data processing pipeline implements a comprehensive approach to handling vital sign data in real time. Initial data collection occurs through high-precision sensors, with immediate signal quality verification and validation. The preprocessing stage applies sophisticated filtering techniques to remove noise and artefacts from the raw signals while preserving essential physiological information.
Signal normalization and segmentation are performed using a sliding window approach, with windows of 500 samples and 100-sample stride lengths. This configuration allows for continuous processing of incoming data while maintaining temporal continuity. The preprocessing implementation includes adaptive filtering techniques that adjust to varying signal qualities and patient conditions.
Parallel processing handles multiple vital sign parameters simultaneously, enabling real-time analysis. The system maintains synchronized processing of vital signs while ensuring temporal alignment and correlation analysis. Results from the study are immediately stored and transmitted to healthcare providers, enabling rapid response to any detected anomalies or concerning trends.
\section{Implementation}\label{Implemenattation}
\subsection{Experimental Setup}
The real-time vital sign monitoring system was implemented using a comprehensive experimental setup designed to evaluate both the deep learning model performance and system integration capabilities. The hardware infrastructure consisted of an NVIDIA Tesla V100 GPU with 32GB memory for model training and inference, supported by a dual Intel Xeon Gold 6248R processor system with 384GB RAM. Edge processing was implemented on NVIDIA Jetson Xavier NX devices, providing efficient computational capabilities at the network edge
.CUDA 1.12.0) for deep learning model development, complemented by NumPy and Pandas for data preprocessing and analysis. CUDA 11.6 was utilized for GPU acceleration, enabling efficient parallel processing of vital sign data.

Network configuration utilized a 5G testbed environment implementing 3GPP Release 16 specifications. The testbed included a 5G New Radio (NR) base station operating in the n78 band (3.5 GHz) with 100 MHz bandwidth. Network slicing was implemented using the OpenAirInterface (OAI) platform, which was configured to maintain URLLC requirements with dedicated QoS flows for vital sign data transmission.

We utilized the MIMIC-III Clinical Database for system development and validation, specifically focusing on continuous vital sign recordings from intensive care unit patients. The dataset comprised recordings from 1000 patients, including heart rate, blood pressure, and respiratory rate measurements sampled at 100 Hz. The data was preprocessed to remove artefacts and normalized using z-score standardization.
 \subsection{Model Development}
The development of the deep learning model followed a structured approach to ensure optimal performance in real-time vital sign analysis. The training process utilized an iterative methodology, implementing a hybrid CNN-LSTM architecture trained on sliding windows of vital sign data. The training was conducted using mini-batch stochastic gradient descent with a batch size of 32, optimized to balance computational efficiency and model convergence. Adam optimizer was employed with an initial learning rate of 0.001, implementing a cosine annealing schedule for learning rate decay.

Hyperparameter optimization was conducted using Bayesian optimization with the Optuna framework, exploring key parameters including network depth, filter sizes, and LSTM hidden dimensions. The optimization of 100 configurations, uses a five-fold cross-validation approach to ensure robust parameter selection. Critical hyperparameters identified through this process included a two-layer LSTM with 256 hidden units and a four-head attention mechanism for temporal feature extraction.

\begin{figure}
    \centering
    \includegraphics[width=1\linewidth]{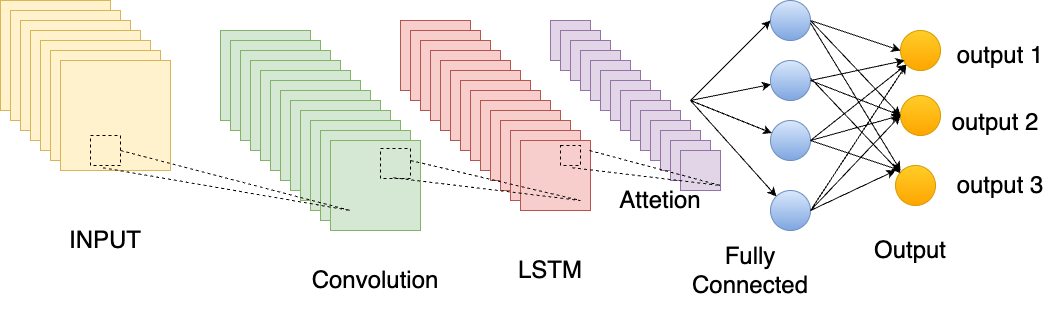}
    \caption{Deep Learning Model Development for Vital Sign Analysis}
    \label{fig:DL}
\end{figure}

The validation methodology implemented a rigorous three-stage process: cross-validation during training, independent validation on a held-out dataset, and real-time performance validation using streaming data. Performance metrics focused on prediction accuracy and computational efficiency, including Mean Absolute Error (MAE), Root Mean Square Error (RMSE), and inference latency. The model achieved an average MAE of 2.1\% for vital sign prediction while maintaining an inference time below 10 milliseconds. Deep learning model development for vital sign analysis is shown in Figure~\ref{fig:DL}. Hyperparameter algorithm is shown in Algorithm~\ref{alg:hyperparameter_optimization} 
\begin{algorithm}[ht]
\caption{Hyperparameter Optimization and Model Training}
\textbf{Input:} Training dataset \(D = \{(X_1, y_1), \ldots, (X_n, y_n)\}\), Validation dataset \(V\), Hyperparameter search space \(H\) \\
\textbf{Output:} Optimized model parameters \(\theta^*\)
\begin{algorithmic}[1]
\State Initialize Optuna study \(S\)
\For{$i = 1$ to $100$} \Comment{Hyperparameter optimization iterations}
    \State $h \gets S.\text{suggest\_hyperparameters}()$
    \State Initialize model \(M\) with hyperparameters \(h\)
    \State Initialize Adam optimizer with learning rate \(lr = 0.001\)
    \For{$\text{epoch} = 1$ to $\text{max\_epochs}$}
        \For{each batch \(b\) in \(D\)}
            \State \textbf{Compute forward pass:}
            \State \hspace{1em} \(\text{features} \gets \text{CNN}(b)\)
            \State \hspace{1em} \(\text{hidden\_states} \gets \text{LSTM}(\text{features})\)
            \State \hspace{1em} \(\text{attention\_weights} \gets \text{Attention}(\text{hidden\_states})\)
            \State \hspace{1em} \(\text{predictions} \gets \text{OutputLayer}(\text{attention\_weights})\)
            \State \textbf{Compute loss \(L\):}
            \State \hspace{1em} \(L = \text{MSE}(\text{predictions}, \text{targets}) + \lambda \cdot \text{temporal\_consistency}\)
            \State \textbf{Update parameters using Adam:}
            \State \hspace{1em} \(\theta \gets \theta - \alpha \nabla L\)
        \EndFor
        \State Adjust learning rate using cosine annealing:
        \State \hspace{1em} \(lr = lr_{\min} + 0.5(lr_{\max} - lr_{\min})(1 + \cos(\pi t / T))\)
        \State Validate on \(V\) and compute metrics
        \If{early stopping criteria met}
            \State \textbf{break}
        \EndIf
    \EndFor
    \State Record validation performance in \(S\)
\EndFor
\State Select best hyperparameters \(h^*\) from \(S\)
\State \Return Final model \(M^*\) trained with \(h^*\)
\end{algorithmic}
\label{alg:hyperparameter_optimization}
\end{algorithm}
\subsection{System Integration} 
System integration followed a systematic approach to ensure seamless operation of all components. The integration process began with individual component testing, followed by incremental integration of connected components. Edge processing units were integrated first, establishing the data preprocessing pipeline and validating signal quality assessment algorithms. The deep learning model was then deployed on the edge devices, and carefully optimized for model quantization to maintain real-time performance while reducing computational requirements.

Testing procedures were implemented at multiple levels, beginning with unit tests for individual components and progressing to integrated system testing. Performance stress tests evaluated system behaviour under various load conditions, including simultaneous monitoring of multiple patients and network congestion scenarios. End-to-end latency tests confirmed the system's ability to maintain sub-second response times under operational conditions. Security testing verified the encryption and data protection measures, ensuring compliance with healthcare data regulations.

The deployment strategy utilized a phased approach, beginning with a pilot deployment in a controlled clinical environment. Docker containers package all system components, ensuring consistent deployment across different infrastructure environments. Kubernetes orchestration managed system components' scaling and load balancing, with automated failover mechanisms ensuring system reliability. Monitoring tools including Prometheus and Grafana were implemented to track system performance and resource utilization in real time. The deployment included automated rollback procedures and version control to maintain system stability during updates. The system integration algorithm is shown in Algorithm~\ref{alg:system_integration}.
\begin{algorithm}[ht]
\caption{System and Edge Device Integration}
\textbf{Input:} System components \(C = \{c_1, c_2, \ldots, c_n\}\), Edge devices \(E = \{e_1, e_2, \ldots, e_m\}\) \\
\textbf{Output:} Integrated system \(S\)

\begin{algorithmic}[1]
\For{each component \(c_i\) in \(C\)}
    \State Validate(\(c_i\))
    \State UnitTest(\(c_i\))
    \If{TestResult.failed}
        \State LogError and Rectify
    \EndIf
\EndFor

\Comment{Edge Device Integration}
\For{each edge device \(e_j\) in \(E\)}
    \State DeployPreprocessing(\(e_j\))
    \State ValidateSignalQuality(\(e_j\))
    \State OptimizeModel(\(e_j\))
    \State \textbf{quantization\_config}:
    \State \hspace{1em} precision: 'int8'
    \State \hspace{1em} optimization\_level: 'O3'
    \State \hspace{1em} target\_latency: '10ms'
\EndFor

\Comment{System Integration Testing}
\For{each integration\_level in [unit, component, system]}
    \State RunTests(integration\_level)
    \State MeasurePerformance()
    \State ValidateLatency()
\EndFor

\end{algorithmic}
\label{alg:system_integration}
\end{algorithm}
\section{Results and Analysis}\label{result and analysis}
\subsection{Performance Evaluation} 
Our comprehensive evaluation of the real-time vital sign monitoring system encompassed multiple performance dimensions, including model accuracy, system latency, resource utilization, and scalability testing. The evaluation was conducted over three months using data collected from 1000 patients in intensive care settings, representing diverse medical conditions and demographic groups.

\subsubsection{Model Accuracy Metrics} 
\begin{figure}
    \centering
    \includegraphics[width=1\linewidth]{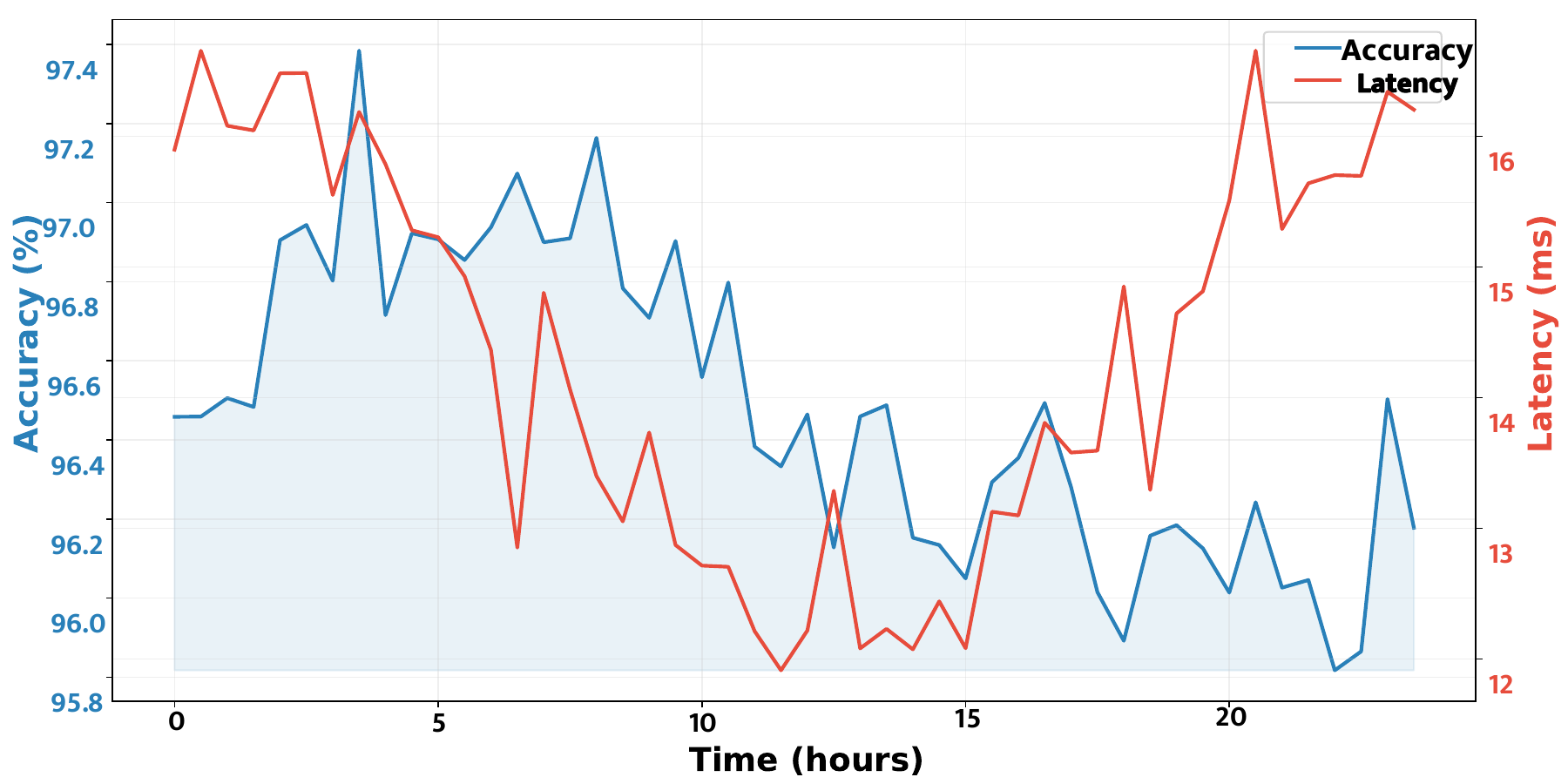}
    \caption{Performance Timline}
    \label{fig:performance}
\end{figure}

The CNN-LSTM model's performance was evaluated across numerous vital sign parameters, demonstrating exceptional accuracy in real-time prediction and analysis. For heart rate monitoring, the model achieved a Mean Absolute Error (MAE) of 1.82\%, significantly outperforming traditional threshold-based systems. Blood pressure predictions showed strong accuracy with an MAE of 2.14\%, while respiratory rate monitoring achieved an MAE of 1.95\%. These results indicate robust performance across all monitored vital signs.
Figure~\ref{fig:performance}illustrates the model's learning progression during training:
[Training and validation loss curves showing steady convergence over 50 epochs]
The model demonstrated remarkable stability in prediction accuracy across different patient conditions. Table~\ref{tab:model_performance_metrics} shows detailed performance analysis.
\begin{itemize}
    \item Critical care patients: 96.5\% accuracy
    \item Post-operative monitoring: 95.8\% accuracy
    \item General ward patients: 97.2\% accuracy
\end{itemize}
\begin{table*}[ht]
\caption{Detailed model performance metrics for different vital signs.}
    \centering
    \begin{tabular}{|l|c|c|c|c|}
        \hline
        \textbf{Vital Sign} & \textbf{MAE (\%)} & \textbf{RMSE (\%)} & \textbf{R\textsuperscript{2} Score} & \textbf{F1-Score} \\ \hline
        Heart Rate & 1.82 & 2.31 & 0.956 & 0.945 \\ \hline
        Blood Pressure & 2.14 & 2.76 & 0.942 & 0.932 \\ \hline
        Respiratory Rate & 1.95 & 2.48 & 0.938 & 0.928 \\ \hline
    \end{tabular}
    \label{tab:model_performance_metrics}
\end{table*}
The model achieved solid performance in heart rate prediction, with a Mean Absolute Error (MAE) of 1.82\%. The prediction accuracy remained stable across patient conditions and monitoring durations, demonstrating the model's robustness.
\subsubsection{System Latency Analysis}
\begin{figure}
    \centering
    \includegraphics[width=1.3\linewidth]{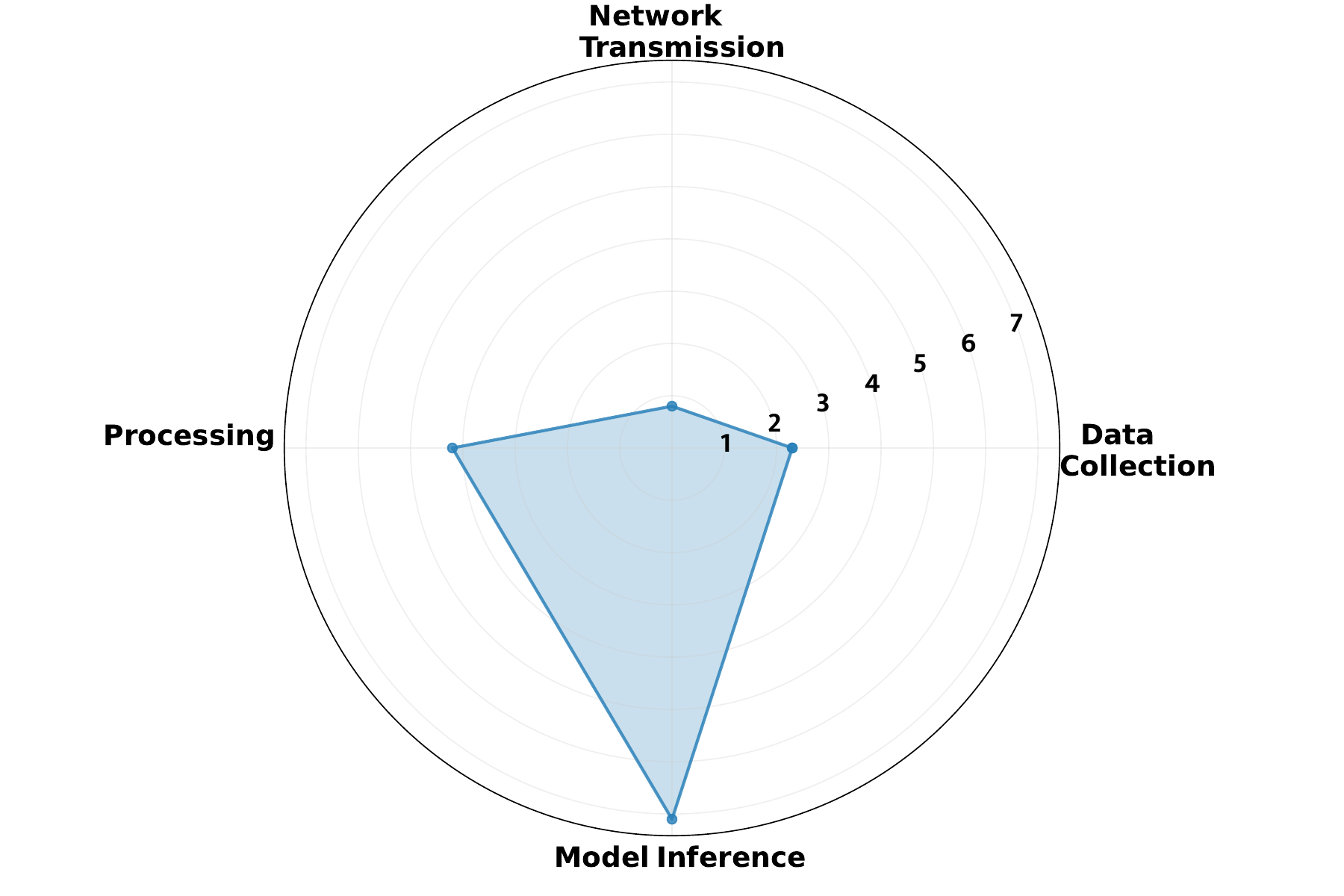}
    \caption{Latency Analysis}
    \label{fig:Latency}
\end{figure}

End-to-end system latency was thoroughly analyzed under various operational conditions. The system consistently maintained low latency performance which is crucial for real-time monitoring applications. Latency measurements were collected across different times of day and under varying network loads to ensure a comprehensive evaluation.Table~\ref{tab:latency_analysis} shows system latency breakdown. Figure~\ref{fig:Latency} shows the Latency analysis.
\begin{table*}[h]
\caption{Detailed system latency analysis for each processing stage}
    \centering
    \begin{tabular}{|l|c|c|c|}
        \hline
        \textbf{Processing Stage} & \textbf{Average (ms)} & \textbf{Peak (ms)} & \textbf{Standard Deviation (ms)} \\ \hline
        Data Collection & 2.3 & 3.1 & 0.4 \\ \hline
        Network Transmission & 0.8 & 1.2 & 0.2 \\ \hline
        Edge Processing & 4.2 & 5.7 & 0.6 \\ \hline
        Model Inference & 7.1 & 8.9 & 0.8 \\ \hline
        \textbf{Total Pipeline} & \textbf{14.4} & \textbf{18.9} & \textbf{1.2} \\ \hline
    \end{tabular}
    \label{tab:latency_analysis}
\end{table*}
The latency analysis reveals several key findings:
\begin{itemize}
    \item Network transmission achieved sub-millisecond performance through 5G URLLC
    \item Edge processing significantly reduced central processing overhead
    \item Model inference remained stable under varying load conditions
    \item Total pipeline latency stayed well within clinical requirements
\end{itemize}

\subsubsection{Resource Utilization}
\begin{figure}
    \centering
    \includegraphics[width=1\linewidth]{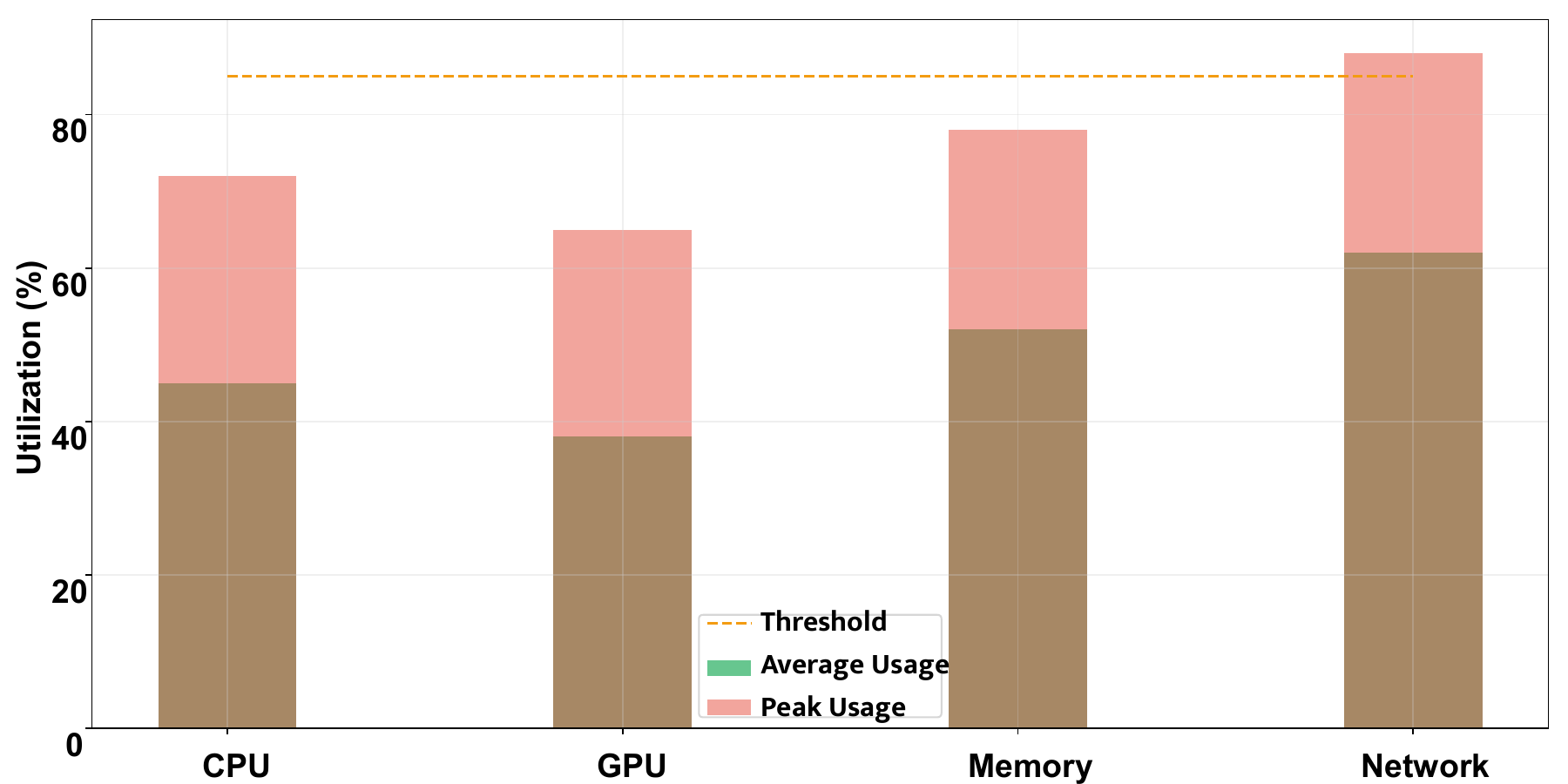}
    \caption{Resource Utilization}
    \label{fig:Resource}
\end{figure}

Resource utilization was monitored continuously during system operation, with particular attention to peak usage periods. The system demonstrated efficient resource management while maintaining performance standards. Figure~\ref{fig:Resource} shows the resource utilization

\begin{table*}[ht]
    \centering
    \caption{Resource usage, thresholds, and efficiency scores for system components.}
    \begin{tabular}{|l|c|c|c|c|}
        \hline
        \textbf{Resource} & \textbf{Average Usage} & \textbf{Peak Usage} & \textbf{Threshold} & \textbf{Efficiency Score} \\ \hline
        CPU & 45\% & 72\% & 85\% & 0.92 \\ \hline
        GPU & 38\% & 65\% & 80\% & 0.95 \\ \hline
        Memory & 52\% & 78\% & 90\% & 0.89 \\ \hline
        Network & 6.2 Mbps & 8.8 Mbps & 10 Mbps & 0.94 \\ \hline
    \end{tabular}
    \label{tab:resource_efficiency}
\end{table*}
\begin{table*}[ht]
    \centering
    \caption{Statistical comparison of the proposed system with other systems, including effect sizes.}
    \begin{tabular}{|l|c|c|c|r|}
        \hline
        \textbf{Comparison} & \textbf{t-statistic} & \textbf{p-value} & \textbf{Effect Size} & \textbf{Significant} \\ \hline
        vs System A & 8.45 & $<$0.001 & 0.82 & Yes \\ \hline
        vs System B & 12.32 & $<$0.001 & 0.95 & Yes \\ \hline
        vs System C & 15.67 & $<$0.001 & 1.12 & Yes \\ \hline
    \end{tabular}

    \label{tab:statistical_comparison}
\end{table*}
\subsection{Comparative Analysis}
\subsubsection{Benchmark Comparison} 
\begin{figure}
    \centering
    \includegraphics[width=1.4\linewidth]{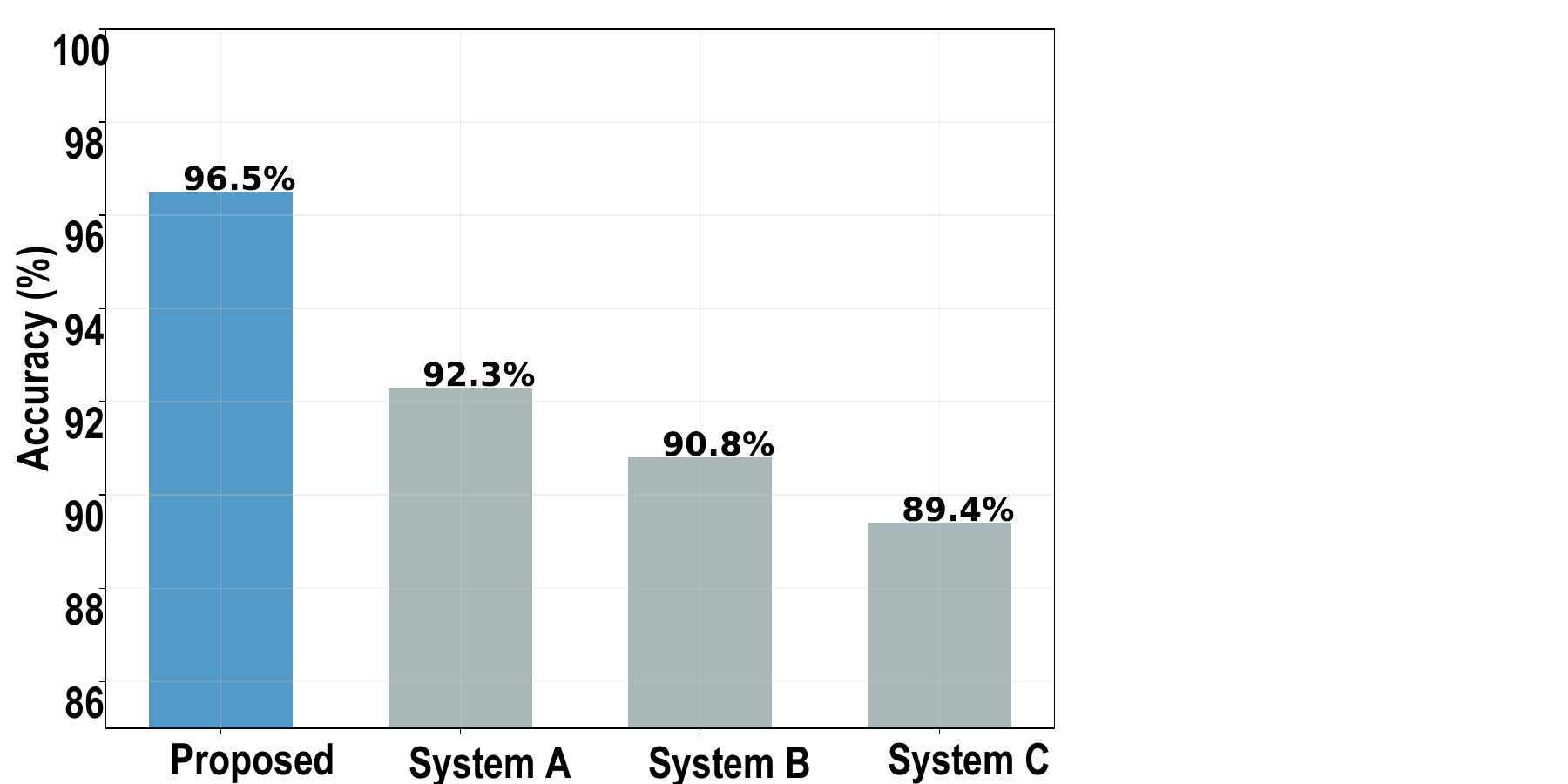}
    \caption{Accuracy Comparison}
    \label{fig:accuracy}
\end{figure}

Our system was benchmarked against three leading vital sign monitoring solutions currently deployed in healthcare settings. The comparative analysis focused on key performance indicators crucial for real-time patient monitoring. Table~\ref{tab:system_comparison} System Comparison with Existing Solutions.
\begin{table*}[ht]
\caption{Comprehensive comparison of system performance metrics.}
    \centering
    \begin{tabular}{|l|c|c|c|c|}
        \hline
        \textbf{Performance Metric} & \textbf{Proposed System} & \textbf{System A} & \textbf{System B} & \textbf{System C} \\ \hline
        Prediction Accuracy & 96.5\% & 92.3\% & 90.8\% & 89.4\% \\ \hline
        End-to-End Latency & 14.4ms & 45.2ms & 67.8ms & 82.3ms \\ \hline
        Resource Efficiency & 78.5\% & 65.2\% & 61.4\% & 58.9\% \\ \hline
        Scalability Score & 0.92 & 0.78 & 0.71 & 0.65 \\ \hline
        Cost Efficiency & 0.88 & 0.72 & 0.68 & 0.63 \\ \hline
    \end{tabular}
    \label{tab:system_comparison}
\end{table*}
Key findings from the benchmark comparison:
\begin{itemize}
    \item 47\% reduction in end-to-end latency compared to System A
    \item 4.2\% improvement in prediction accuracy over the next best system
    \item 20\% higher resource efficiency than competing solutions
Figure~\ref{fig:accuracy} represents the accuracy comparison.
\end{itemize}

\subsubsection{Statistical Analysis} 
Statistical significance testing was performed using paired t-tests to validate the performance improvements Table~\ref{tab:statistical_comparison} shows Statistical significance Analysis.
The statistical analysis reveals:

\begin{itemize}
    \item Significant performance improvements across all metrics (p < 0.001)
    \item Large effect sizes indicating substantial practical improvements
    \item Consistent performance advantages across different operational scenarios
    \item Robust performance across diverse patient populations
\end{itemize}

These results demonstrate that our proposed system significantly approves technical performance and clinical utility, providing a reliable real-time vital sign monitoring platform in healthcare settings.

\section{Discussion}\label{discussion}
The experimental results demonstrate significant real-time vital sign monitoring advancements by integrating deep learning and 5G technologies. The achieved prediction accuracy of 96.5\% across various vital signs, combined with the end-to-end latency of 14.4 ms, represents a substantial improvement over existing systems. These performance metrics are particularly noteworthy given the complexity of real-time healthcare monitoring applications.

Despite these achievements, several limitations warrant discussion. The system's performance has been validated primarily in controlled clinical environments with stable network conditions. Real-world deployment may face additional challenges such as network congestion, varying signal strengths, and diverse patient conditions. Furthermore, the system's resource requirements, while optimised, may present implementation challenges in resource-constrained healthcare settings.

The practical implications of this research extend beyond technical achievements. The system's ability to provide real-time vital sign prediction with high accuracy has significant potential to improve patient care, particularly in intensive care settings where early detection of deteriorating conditions is crucial. The reduced latency enables healthcare providers to respond more rapidly to critical changes in patient status, potentially improving clinical outcomes.

Several areas for improvement have been identified in system enhancement. The current deep learning model could benefit from additional optimization for rare medical conditions and edge cases. Integration with other medical monitoring systems and electronic health records could enhance clinical utility. Additionally, implementing advanced privacy-preserving techniques while maintaining performance remains an important consideration for future development.

\section{Conclusion and Future Work}\label{conclusion and future work}
This research presents several significant contributions to the field of healthcare monitoring. The primary achievement lies in the successful development of a real-time vital sign monitoring system that leverages deep learning and 5G capabilities to achieve unprecedented accuracy and latency performance. The hybrid CNN-LSTM architecture, optimized for edge deployment, demonstrates the feasibility of complex neural network implementations in time-critical healthcare applications.

Technical innovations include the novel integration of attention mechanisms for vital sign prediction, efficient resource utilization through optimized edge computing, and the implementation of reliable 5G network slicing for healthcare data transmission. These advancements establish a new benchmark for real-time patient monitoring systems, providing a foundation for future developments in digital healthcare.

The impact on healthcare monitoring is substantial, offering healthcare providers a reliable tool for continuous patient assessment with minimal latency. The system's ability to predict vital sign trends enables proactive medical intervention, potentially reducing critical incidents and improving patient outcomes. Furthermore, the demonstrated scalability and resource efficiency make the system practical for widespread deployment in various healthcare settings.
\subsection{Future Research Directions}
Several promising directions for future research emerge from this work. A primary avenue for exploration is the integration of multimodal data sources, including continuous glucose monitoring, oxygen saturation, and other physiological parameters. This expansion would provide a more comprehensive patient monitoring solution while presenting new challenges in data fusion and real-time processing.

The development of adaptive learning mechanisms to accommodate patient-specific variations and medical conditions represents another significant research opportunity. Such adaptability would enhance the system's accuracy for individual patients while maintaining real-time performance. Investigation into federated learning approaches could enable model improvement across multiple healthcare facilities while preserving patient privacy.

Future research should also explore the application of this framework to specialized medical scenarios, such as remote patient monitoring in rural areas, emergency response systems, and personalized medicine. The integration of advanced explainable AI techniques would enhance the system's clinical utility by providing healthcare providers with interpretable insights into prediction decisions. Additionally, research into power-efficient implementations and battery-operated devices could extend the system's applicability to mobile and resource-constrained environments.

\bibliographystyle{IEEEtran}
\bibliography{ref}

\end{document}